# The Challenge of Unconventional Superconductivity

Michael R. Norman

Materials Science Division, Argonne National Laboratory, Argonne, IL  60439
e-mail:  norman@anl.gov

**During the past few decades, several new classes of superconductors have been discovered. Most of these do not appear to be related to traditional superconductors. As a consequence, it is felt by many that for these materials, superconductivity arises from a different source than the electron-ion interactions that are at the heart of conventional superconductivity. Developing a rigorous theory for any of these classes of materials has proven to be a difficult challenge, and will continue to be one of the major problems in physics in the decades to come.**

Superconductivity is an exotic state of matter that has intrigued scientists ever since its discovery in mercury in 1911 (*1*).  It is sobering to realize that after one hundred years, there are whole classes of superconducting materials that we still do not fully understand.  Even for the first ('conventional') superconductors, it took half a century to develop a theory, starting with the ground-breaking work of Cooper in 1956 (*2*) and Bardeen, Cooper and Schrieffer (BCS) in 1957 (*3*). These ideas rapidly developed into a rigorous theory several years later (*4*), since for electron-ion interactions, a controlled many-body perturbation expansion is possible that greatly reduced the complexity of the problem (*5*). Unfortunately, we do not have this luxury for the later discovered unconventional superconductors, which is why developing rigorous theories for these materials is extremely challenging.

## Conventional vs. Unconventional Superconductors

To appreciate these issues, we need to first understand what superconductors are all about, and how unconventional ones differ from their more conventional counterparts.  Superconductors are not only perfect conductors (their electrical resistance drops precipitously to zero below a transition temperature $T_c$), but also exhibit the so-called Meissner effect (*6*), where they expel magnetic fields.  As noted by Fritz London (*7*), this implies that electrons in superconductors behave in a collective manner. Bosons, which have integer values of a fundamental property known as 'spin', can behave in this fashion, whereas electrons, which are fermions that have half-integer spins, typically do not. This apparent contradiction was resolved by Leon Cooper in 1956 (*2*) who demonstrated that the presence of even an arbitrarily small attractive interaction between the electrons in a solid causes the electrons to form pairs.  Because these 'Cooper pairs' behave as effective bosons, they can form something analogous to a Bose-Einstein condensate. Rather than being real-space molecules, however, Cooper pairs consist of electrons in time-reversed momentum states and consequently have zero center-of-mass momentum.  Because a pair of identical fermions is antisymmetric with respect to the exchange of one fermion with another, the spin and spatial components of the Cooper pair wavefunction must have opposite exchange symmetries. Thus these pair states are either spin singlets with an even parity spatial component, or spin triplets with odd parity.  The spin singlet pair state with an isotropic spatial component (s-wave) turns out to be the one realized in conventional superconductors (*3*).

Despite the fact that electrons repel each other because of the Coulomb force, at low energies there can be an effective attraction resulting from the electron-ion interaction. To understand this, note that a metal is formed by mobile electrons detaching themselves from the atoms that form the crystalline lattice (these atoms then become positive ions). Such a mobile electron attracts the surrounding ions because of their opposite charge. When this electron moves, a positive ionic distortion is left in its wake. This attracts a second electron, leading to a net attraction between the electrons. This mechanism works because the ion dynamics is slow compared to the electrons, a consequence of the fact that the ions are much heavier than the electrons. However, the interaction at shorter times becomes repulsive because of the Coulomb interaction between the electrons; this retardation is what is responsible for limiting $T_c$ (*8*). Up until the discovery of cuprates, the highest known $T_c$ was only 23K.

The resulting wavefunction for the pairs turns out to be peaked at zero separation of the electrons – that is, an s-wave state. By an `unconventional' superconductor, we mean one that is not s-wave. Typically, this means that the pair wavefunction, also known as the order parameter, is not uniform in momentum space, though there can be exceptions, as may be the case for the new iron-based superconductors.

### $^3$He – the First Unconventional Superfluid

The first unconventional material was not a superconductor, however, but rather superfluid $^3$He (*9*). Because the atoms are neutral objects in this case, there is no analogue of the electron-ion interaction mentioned above. In fact, the helium atoms behave as hard core objects, which acts to suppress s-wave pairing. One of the first proposals to explain the pairing was based on van der Waals attraction between the atoms at larger separations (*10*) resulting in the formation of d-wave pairs. This pair state suppresses the influence of the hard core because it has a node for zero separation of the atoms, and the d-wave state optimizes the separation of the atoms to take advantage of the van der Waals attraction. A p-wave state, which has a node as well, was also proposed, but because of fermion antisymmetry, this odd parity state is associated with spin triplet pairs. In this case, the attraction was speculated to be a result of exchange forces (*11*); $^3$He was thought to be nearly ferromagnetic, so an atom would prefer to align its spin with its pair partner. When superfluidity was seen in $^3$He several years later, it was soon realized that it was indeed a consequence of p-wave pairing (*12*). Two different superfluid phases, known as A and B, exist below a few mK, which were also explained based on exchange forces (*13*). Given the simplicity of the liquid state, the pair interaction was eventually quantified using known normal state interaction (Landau) parameters. This analysis revealed that many factors contribute to the pair interaction, including density, spin, and transverse current interactions (*12*). This provides a cautionary tale that it is dangerous to claim that one mechanism is the sole cause of pairing in unconventional superconductors.

### Heavy Fermion Superconductors

Of course, neutral atoms are not the same as charged electrons. Surprisingly, superconductivity with a $T_c$ of 0.5K was discovered in 1979 in $CeCu_2Si_2$ (Fig. 1A) by Steglich's group (*14*), and then in several uranium alloys such as $UPt_3$ and $UBe_{13}$ (*15*) a few years later. These materials contain magnetic 4f and 5f ions, which based on previous experience would have been incompatible with superconductivity. After all, magnetic impurities are well known sources of pair breaking in conventional superconductors (*16*). But these materials are more intriguing in that the ions exhibit the Kondo effect. In such systems, the mobile conduction electrons have a tendency to form a bound resonance with the localized f electrons of the magnetic ions (*17*). When these magnetic ions form a regular lattice, this `Kondo' lattice is characterized by an electronic specific heat coefficient of order 1000 times larger than conventional metals like copper, and a correspondingly large spin susceptibility. This `heavy' Fermi liquid can form a number of ordered states, including magnetic order, and, intriguingly, unconventional superconductivity (*15*).

Given some similarities with $^3$He, where the liquid phase is near a solid phase with magnetic order, there were initial speculations that these heavy fermion metals were p-wave superconductors as well (*18*). But subsequent work has indicated that this is a complex problem (*19*). First, the concept of s-wave, p-wave, etc., has to be taken with a grain of salt because of the presence of a crystalline lattice that breaks translational symmetry. Second, not only are multiple f orbitals involved, there are multiple conduction electron orbitals as well. Finally, spin-orbit effects are large for Ce and U ions, and play a qualitatively different role than in light $^3$He atoms. In fact, although it has been over thirty years since their discovery, the actual symmetry of the Cooper pairs of any heavy fermion superconductor has yet to be unambiguously determined. Perhaps the closest we have come is UPt$_3$ (*20*). This material has several superconducting phases (Fig. 2A), like $^3$He. Because of the hexagonal symmetry of the lattice and the fact that the spin and orbital angular momentum of the pairs are linked because of spin-orbit coupling, this implies that either two different superconducting states have nearly identical transition temperatures, or that the order parameter is doubly degenerate. The latter is more likely, as it is thought that the small temperature range separating the two phases at zero magnetic field is a result of a weak lifting of the hexagonal symmetry of the lattice caused by the presence of small magnetic moments on the uranium ions. A variety of thermodynamic data indicates that the order parameter probably vanishes along lines that are perpendicular to the c axis of the crystal. This restricts the order parameter symmetry to be either $E_{1g}$ (`d-wave') or $E_{2u}$ (`f-wave'). Very recently, measurements have been carried out that are sensitive to the phase of the order parameter (*21*). They indicate the order parameter behaves as $e^{2i\phi}$ where $\phi$ is the azimuthal angle within the hexagonal plane. If a line of nodes is indeed present, this rules out $E_{1g}$ in favor of $E_{2u}$. Such an f-wave order parameter is indeed an exotic beast (Fig. 2B). But it is highly doubtful that the actual order parameter is so simple – UPt$_3$ has a complex Fermi surface (which separates occupied from unoccupied states) formed from five different energy bands, and the spin-orbit coupling is so strong that even the single particle states are best characterized by states of total angular momentum J.

Less is known about other heavy fermion superconductors. Much recent work has gone into the so-called 115 series, with a formula unit CeXIn$_5$ where X is a transition metal ion (Fig. 1B). Available data are consistent with a non-degenerate order parameter that is singlet in nature, leading to the speculation of d-wave pairs. But to date, no phase sensitive measurements have been performed. Of perhaps greater interest are the plutonium analogues, one of which has a T$_c$ of 18K, almost an order of magnitude higher than previously known heavy fermion superconductors (*22*). Less is known about this material given the challenges of working with plutonium, but its discovery indicates that perhaps even more dramatic examples await us in the future. Already there are heavy fermion superconductors, such as $UGe_2$ and $URhGe$, which are simultaneously ferromagnetic and superconducting. And perhaps related to these systems is $Sr_2RuO_4$ (23), a multi-band layered transition metal oxide that appears to be a p-wave superconductor, though the exact nature of the order parameter is still being debated. Available phase sensitive measurements are certainly consistent with a p-wave state (24).

A fundamental difference from $^3$He, though, is that most heavy fermion superconductors are actually nearly antiferromagnetic, and in some cases, most dramatically in the Ce 115 series, superconductivity and antiferromagnetism co-exist (Fig. 1). This was realized back in 1985 when strong antiferromagnetic spin correlations were seen by neutron scattering in $UPt_3$ (25), leading to the publication of three theoretical papers advocating that antiferromagnetic spin fluctuations were the source of d-wave superconductivity (26-28). Superconductivity seems to be maximal at a point where magnetism disappears (Fig. 1), something first appreciated in 1998 when superconductivity was discovered under pressure in $CePd_2Si_2$ and $CeIn_3$ (29). A phase transition suppressed to zero temperature is known as a quantum phase transition (30), and it is thought that critical fluctuations associated with this quantum critical point could be the source of pairing (29).

**Cuprates**

In the same year the above mentioned spin fluctuation papers were published, a small group from an IBM lab in Zurich made a startling discovery – superconductivity near 40K in the layered cuprate $La_{2-x}Ba_xCuO_4$ (31). At first, this result did not attract much attention (in the past, there had been a number of claims of USOs – unidentified superconducting objects). But after its reproduction by several groups, a flurry of activity was unleashed, leading shortly to the discovery of superconductivity above 90K in $YBa_2Cu_3O_7$ (32). The technological implications were profound, given the breaking of the `liquid air barrier'. Surprisingly, this class of materials violated most if not all of the empirical search rules set down by Bernd Matthias, based on the previous record high $T_c$ materials which were cubic transition metal alloys – as opposed to cuprates, which are obtained by doping carriers into a parent material that is an insulating magnetic oxide. Not surprisingly, theorists speculated that the solution for the cuprate puzzle was a 2D variant of the d-wave superconductivity mentioned above in the heavy fermion context, $d_{x^2-y^2}$ symmetry (33).

But before this, a very different theory appeared that for better or worse would change the face of physics (34). The Nobel Laureate, Philip Anderson, proposed instead that cuprates would exhibit a novel phase of matter where the spins formed a liquid of singlets – the so-called RVB (resonating valence bond) state based on previous work he had done in the 1970s on frustrated magnets. The name RVB was motivated by the classic work of Linus Pauling on benzene rings, where the carbon bonds fluctuate between single and double bonds. Anderson argued that such an RVB state was the consequence of several unique properties of cuprates – the materials are quasi-two dimensional, the copper ions have spin ½, and the parent phase is a Mott insulator, that is, a state with an odd number of electrons per unit cell that is insulating because of many-body correlations. These effects he speculated would act to melt the expected antiferromagnetic (Néel) lattice into this spin liquid phase. Upon carrier doping, these singlets would become charged, resulting in a superconducting state. Although the original proposal was for a uniform

RVB (s-wave) state, subsequent work found that the free energy was actually minimized for a d-wave state (*35*). Although undoped cuprates were soon found to form a Néel lattice (though with a reduced moment), a few percent of doped holes was sufficient to destroy this state (Fig. 3A).

What is unquestionable is that the exchange interaction $J$ for cuprates is very large, of order 1400K, and as such is an attractive source for pairing. This became very relevant in the mid 1990s, when it was shown by phase sensitive tunneling (*36*) that the pairing state was indeed d-wave (Fig. 3B). But such a large $J$ is also relevant for the more traditional spin fluctuation based approaches, and there is currently much debate about which of these two approaches, RVB (*37*) versus spin fluctuations (*38*), is the more appropriate. The lack of resolution of this debate is connected to the fact that for electronic only models, we do not have a controlled perturbation expansion to work with as we do for the electron-ion interactions underlying conventional superconductors. Moreover, cuprates are complex systems, with a variety of important interactions, including electron-ion. This has become increasingly obvious in attempts to explain their phase diagram (Fig. 3A). After the Néel order is destroyed by doping, there are four apparent regions of the phase diagram: (1) a pseudogap phase where an energy gap is present, (2) a strange metal phase characterized by a resistivity linear in temperature, (3) a Fermi liquid phase with largely normal transport properties, and (4) a d-wave superconducting phase. In the RVB approach, the pseudogap is a spin gap phase resulting from spin singlet formation, whereas in the spin fluctuation approach, it is a fluctuating version of the Néel phase. But experiments now indicate an intriguing variety of phenomena associated with the pseudogap phase, including nematic (*39*) correlations (where the $C_4$ rotational symmetry of the square lattice is spontaneously broken), and a novel form of magnetism, either coming from orbital currents or antiferromagnetism, which is associated with the oxygen sites in the $CuO_2$ unit cell of the cuprates (*40*). At lower temperatures, charge density wave, spin density wave, and superconducting correlations become apparent in a variety of measurements. From this very complicated soup, high temperature superconductivity arises. Because superconductivity is created from the normal state, these phenomena must be understood before we will ever have a true understanding of the origin of high temperature cuprate superconductivity. Several authors have pointed out the similarities of the phase diagram of cuprates (Fig. 3A) with that of heavy fermions (Fig. 1). Based on this, it has been proposed that superconductivity is mediated by quantum critical fluctuations, but again, the nature of the purported quantum critical point (which in the case of the cuprates is `hidden' under the superconducting dome) is being actively debated. Is it associated with antiferromagnetism, charge density wave, spin density wave, nematic, orbital currents, or a combination thereof?

**Organic superconductors**

If the physics of doped Mott insulators is indeed the key to cuprates (*37*), this will also have relevance to organic superconductors (*41*). These materials were discovered well before the cuprates, and were an equal surprise to the community, Bernd Matthias once quipping "there aren't any!" Well, they are indeed real, and the 2D variety has a phase diagram intriguingly similar to their cuprate counterpart (Fig. 4a). In the BEDT-TTF salts, one has a lattice of molecular dimers, with one spin ½ degree of freedom per dimer. These are arranged in a triangular fashion, which has a tendency to suppress magnetic order due to frustration. It was indeed such a lattice that was the motivation for the original RVB idea. These materials at ambient pressure typically have a Mott insulating ground state, becoming superconducting under pressure (*42*). Available evidence indicates a d-wave state, but as with other unconventional superconductors except cuprates, the true pairing symmetry has yet to be unambiguously determined. Much of the recent attention regarding these materials has been devoted to those compounds which seem to exhibit a spin liquid ground state in the Mott phase, as well as to the question of whether a Fermi surface of spin degrees of freedom (a so-called spinon Fermi surface) is indeed realized (*43*) as originally proposed by Anderson for the cuprates (*34*). As with the cuprates, there has been an interesting debate regarding RVB versus spin fluctuation approaches for the pairing mechanism.

**Iron-based Superconductors**

This brings us to the newly discovered iron-based pnictide and related superconductors, which will be covered in a companion article. Although involving iron rather than copper, and arsenic rather than oxygen, there are enough similarities to cuprates for these debates to again occur for this class of compounds (*44*). Unlike the cuprates, it appears that all five of the 3d orbitals of the iron are involved in the electronic structure near the Fermi energy. Although the undoped material is also antiferromagnetic, unlike in the cuprates, it is metallic. With doping, the magnetic state is suppressed, and a high temperature superconducting phase appears (Fig. 4B). It has been speculated that the order parameter is a so-called $s_\pm$ state, where the Fermi surfaces around the $\Gamma$ point of the Brillouin zone have an order parameter with one sign, and those around the M point of the zone the opposite sign (*45*). Like the cuprates, and what has been speculated for the heavy fermions, this order parameter satisfies the condition that it changes sign under translation by the magnetic ordering vector of the parent phase. Such a state tends to remove the detrimental effects of the on-site Coulomb repulsion between the electrons, and naturally appears in spin fluctuation models. The jury, though, is still out on this question. For sure, this order parameter is consistent with what is known from angle resolved photoemission, which shows rather isotropic energy gaps around each Fermi surface (*46*), but thermal conductivity (*47*) and other measurements indicate for certain dopings, gaps that appear to be d-wave-like in nature. And although the correlations appear to be weaker than in the cuprates due to the multi-orbital nature of the electronic structure (the electrons can avoid one another by occupying different d orbitals), there has been a rather heated debate whether the magnetism is itinerant like the spin density wave state in chromium, or more localized as in the case of the cuprates. Interestingly, strong nematic effects similar to what have been observed in the cuprates have been seen in these materials as well (*48*).

**Theory and Outlook**

From a theoretical point of view, what has become increasingly obvious is that unconventional superconductivity is a very tough problem. Even for the simple case of one d orbital with an on-site Coulomb repulsion, which has been considered by many to be the minimal description for cuprates, we do not know whether this model is indeed superconducting. Quantum Monte Carlo simulations of this model have given conflicting results, the issue being the infamous fermion sign problem that plagues such simulations. Even if it turns out to be superconducting, there are many who feel this model is not sufficient – for instance, because of the rather large electron-ion effects that exist even for cuprates (*49*), or because a three band model may be necessary to describe the physics (*50*). For sure, multi-band models are unavoidable in many cases, most prominently for heavy fermions, ruthenates, and pnictides. And unlike in conventional superconductors, in electronic only mechanisms, there is no really controlled perturbation theory to work with. Theorists have attempted to get around this by suggesting that an effective expansion parameter might exist, exploiting the fact that the collective spin fluctuations are "slow" compared to individual fermion degrees of freedom. On the other hand, it is well known that some higher order terms are as large as the lowest order term in such theories (*51*). Even in the RVB approach, most of the work has been done at an effective mean field level reminiscent of BCS theory. Attempts to go beyond this by considering `gauge' fluctuations associated with constraints such as no double occupancy on a given copper site have met with limited success (*37*). In both spin fluctuation and in the RVB "gauge" approaches, large N expansions (where N is the degeneracy of the electronic states) have been attempted motivated by earlier work in heavy fermion materials. But even in heavy fermions where the orbital degeneracy of the f electrons is large, N is typically 2 in the low energy sector because of crystal field splittings, so the relevance of these approaches is controversial. Even the string theorists have gotten into the act, suggesting that the AdS/CFT (anti de Sitter/conformal field theory) methods they have used to study certain strong coupling gauge theories may be relevant to condensed matter systems like cuprates (*52*). Certainly, in the coming years, there will be increasing attention given to constructing rigorous non-perturbative strong coupling theories in the hope that we can "solve" the problem of unconventional superconductivity, at least to most people's satisfaction.

In regards to the materials themselves, one might ask "what's next?" Since the discovery of cuprates, several new families of superconductors have been discovered. $MgB_2$ appears to be a conventional superconductor, and its properties were explained rather quickly by the standard strong-coupling approach for electron-ion interactions (*53*). It is revealing to note that this simple material was missed by both the experimentalists and theorists for such a long period of time. Alkali-doped buckyballs also have a substantial $T_c$ near 40K, and although at first sight appear to be conventional, there is increasing evidence that Mott physics may be involved here as well (*54*). And the discovery of pnictides again took the community by surprise. Because of this, there is no doubt that new classes of superconductors await our discovery. Whether current beliefs will aid us in this quest remains to be seen. The discovery of large values of the exchange constant *J* in iridium oxides (*55*) has led a few of us to propose that these materials might also be high temperature superconductors. But attempts to dope these materials have so far not led to a superconducting phase. There are also a number of layered nitrides that have a comparatively high $T_c$ near 26K (*56*). Perhaps different variants are out there that will take us into the $T_c$ range of cuprates. Regardless, what is apparent is that such discoveries will only occur with a proper investment in materials synthesis, guided by a good intuition of where to look. If this occurs, the future will indeed be bright.

**References and Notes**

1. H. Kammerlingh Onnes, *Leiden Comm.* **120b**, **122b**, **124c** (1911).
2. L. N. Cooper, Bound Electron Pairs in a Degenerate Fermi Gas. *Phys. Rev.* **104**, 1189 (1956).
3. J. Bardeen, L. N. Cooper, J. R. Schrieffer, Theory of Superconductivity. *Phys. Rev.* **108**, 1175 (1957).
4. J. R. Schrieffer, D. J. Scalapino, J. W. Wilkins, Effective Tunneling Density of States in Superconductors. *Phys. Rev. Lett.* **10**, 336 (1963).
5. G. M. Eliashberg, Interactions between Electrons and Lattice Vibrations in a Superconductor. *Sov. Phys. JETP* **11**, 696 (1960).
6. W. Meissner and R. Ochsenfeld, A new effect concerning the onset of superconductivity. *Naturwissenschaften* **21**, 787 (1933).
7. F. London, *Superfluids*, vol. I (Wiley, New York, 1950).
8. P. Morel and P. W. Anderson, Calculation of the Superconducting State Parameters with Retarded Electron-Phonon Interaction. *Phys. Rev.* **125**, 1263 (1962).
9. D. D. Osheroff, R. C. Richardson, D. M. Lee, Evidence for a New Phase of Solid $He^3$. *Phys. Rev. Lett.* **28**, 885 (1972).
10. V. J. Emery, A. M. Sessler, Possible Phase Transition in Liquid $He^3$. *Phys. Rev.* **119**, 43 (1960).
11. D. Fay, A. Layzer, Superfluidity of Low-Density Fermion Systems. *Phys. Rev. Lett.* **20**, 187 (1968).
12. A. J. Leggett, A theoretical description of the new phases of liquid $^3$He. *Rev. Mod. Phys.* **47**, 331 (1975).
13. P. W. Anderson, W. F. Brinkman, Anisotropic Superfluidity in $^3$He: A Possible Interpretation of Its Stability as a Spin-Fluctuation Effect. *Phys. Rev. Lett.* **30**, 1108 (1973).
14. F. Steglich *et al.*, Superconductivity in the Presence of Strong Pauli Paramagnetism: $CeCu_2Si_2$. *Phys. Rev. Lett.* **43**, 1892 (1979).
15. G. R. Stewart, Heavy-fermion systems. *Rev. Mod. Phys.* **56**, 755 (1984).
16. A. A. Abrikosov, L. P. Gor'kov, Contribution to the Theory of Superconducting Alloys with Paramagnetic Impurities. *Sov. Phys. JETP* **12**, 1243 (1961).
17. K. G. Wilson, The renormalization group: Critical phenomena and the Kondo problem. *Rev. Mod. Phys.* **47**, 773 (1975).
18. P. A. Lee, T. M. Rice, J. W. Serene, L. J. Sham, J. W. Wilkins, Theories of Heavy-Electron Systems. *Comments Cond. Matter Phys.* **12**, 99 (1986).
19. C. Pfleiderer, Superconducting phases of f-electron compounds. *Rev. Mod. Phys.* **81**, 1551 (2009).
20. R. Joynt, L. Taillefer, Superconducting phases of f-electron compounds. *Rev. Mod. Phys.* **74**, 235 (2002).
21. J. D. Strand *et al.*, The Transition Between Real and Complex Superconducting Order Parameter Phases in $UPt_3$. *Science* **328**, 1368 (2010).
22. J. L. Sarrao *et al.*, Plutonium-based superconductivity with a transition temperature above 18 K. *Nature* **420**, 297 (2002).
23. Y. Maeno *et al.*, Superconductivity in a layered perovskite without copper. *Nature* **372**, 532 (1994).
24. K. D. Nelson, Z. Q. Mao, Y. Maeno, Y. Liu, Odd-Parity Superconductivity in $Sr_2RuO_4$. *Nature* **306**, 1151 (2004).

25. G. Aeppli, A. Goldman, G. Shirane, E. Bucher, M.-Ch. Lux-Steiner, Development of antiferromagnetic correlations in the heavy-fermion system $UPt_3$. *Phys. Rev. Lett.* **58**, 808 (1987).
26. K. Miyake, S. Schmitt-Rink, C. M. Varma, Spin-fluctuation-mediated even-parity pairing in heavy-fermion superconductors. *Phys. Rev. B* **34**, 6554 (1986).
27. M. T. Beal-Monod, C. Bourbonnais, V. J. Emery, Possible superconductivity in nearly antiferromagnetic itinerant fermion systems. *Phys. Rev. B* **34**, 7716 (1986).
28. D. J. Scalapino, E. Loh, J. E. Hirsch, d-wave pairing near a spin-density-wave instability. *Phys. Rev. B* **34**, 8190 (1986).
29. N. D. Mathur *et al.*, Magnetically mediated superconductivity in heavy fermion compounds. *Nature* **394**, 39 (1998).
30. Q. Si, F. Steglich, Heavy Fermions and Quantum Phase Transitions. *Science* **329**, 1161 (2010).
31. J. G. Bednorz, K. A. Muller, Possible High $T_c$ Superconductivity in the Ba-La-Cu-O System. *Zeit. Phys. B* **64**, 189 (1986).
32. M. K. Wu *et al.*, Superconductivity at 93 K in a new mixed-phase Y-Ba-Cu-O compound system at ambient pressure. *Phys. Rev. Lett.* **58**, 908 (1987).
33. N. E. Bickers, D. J. Scalapino, R. T. Scaletar, CDW and SDW Mediated Pairing Interactions. *Intl. J. Mod. Phys. B* **1**, 687 (1987).
34. P. W. Anderson, The Resonating Valence Bond State in $La_2CuO_4$ and Superconductivity. *Science* **235**, 1196 (1987).
35. G. Kotliar, J. Liu, Superexchange mechanism and d-wave superconductivity. *Phys. Rev. B* **38**, 5142 (1988).
36. C. C. Tsuei, J. R. Kirtley, Pairing symmetry in cuprate superconductors. *Rev. Mod. Phys.* **72**, 969 (2000).
37. P. A. Lee, N. Nagaosa, X.-G. Wen, Doping a Mott insulator: Physics of high-temperature superconductivity. *Rev. Mod. Phys.* **78**, 17 (2006).
38. T. A. Maier, D. Poilblanc, D. J. Scalapino, Dynamics of the Pairing Interaction in the Hubbard and t-J Models of High-Temperature Superconductors. *Phys. Rev. Lett.* **100**, 237001 (2008).
39. M. J. Lawler *et al.*, Intra-unit-cell electronic nematicity of the high-$T_c$ copper-oxide pseudogap states. *Nature* **466**, 347 (2010).
40. B. Fauque *et al.*, Magnetic Order in the Pseudogap Phase of High-$T_c$ Superconductors. *Phys. Rev. Lett.* **96**, 197001 (2006).
41. B. J. Powell, R. H. McKenzie, Strong electronic correlations in superconducting organic charge transfer salts. *J. Phys. Cond. Matter* **18**, R827 (2006).
42. F. Kagawa, K. Miyagawa, K. Kanoda, Unconventional critical behaviour in a quasi-two-dimensional organic conductor. *Nature* **436**, 534 (2005).
43. P. A. Lee, An End to the Drought of Quantum Spin Liquids. *Science* **321**, 1306 (2008).
44. M. R. Norman, High-temperature superconductivity in the iron pnictides. *Physics* **1**, 21 (2008).
45. I. I. Mazin, D. J. Singh, M. D. Johannes, M. H. Du, Unconventional Superconductivity with a Sign Reversal in the Order Parameter of $LaFeAsO_{1-x}F_x$. *Phys. Rev. Lett.* **101**, 057003 (2008).
46. H. Ding *et al.*, Observation of Fermi-surface–dependent nodeless superconducting gaps in $Ba_{0.6}K_{0.4}Fe_2As_2$. *EPL* **83**, 47001 (2008).

47. M. A. Tanatar *et al.*, Doping Dependence of Heat Transport in the Iron-Arsenide Superconductor Ba(Fe$_{1-x}$Co$_x$)$_2$As$_2$: From Isotropic to a Strongly k-Dependent Gap Structure. *Phys. Rev. Lett.* **104**, 067002 (2010).
48. J.-H. Chu *et al.*, In-Plane Resistivity Anisotropy in an Underdoped Iron Arsenide Superconductor. *Science* **329**, 824 (2010).
49. D. Reznik *et al.*, Electron–phonon coupling reflecting dynamic charge inhomogeneity in copper oxide superconductors. *Nature* **440**, 1170 (2006).
50. C. M. Varma, Theory of the pseudogap state of the cuprates. *Phys. Rev. B* **73**, 155113 (2006).
51. J. A. Hertz, K. Levin, M. T. Beal-Monod, Absence of a Migdal theorem for paramagnons and its implications for superfluid He$^3$. *Solid State Comm.* **18**, 803 (1976).
52. T. Faulkner, N. Iqbal, H. Liu, J. McGreevy, D. Vegh, Strange Metal Transport Realized by Gauge/Gravity Duality. *Science* **329**, 1043 (2010).
53. J. Kortus, I. I. Mazin, I. D. Belashchenko, V. P. Antropov, L. L. Boyer, Superconductivity of Metallic Boron in MgB$_2$. *Phys. Rev. Lett.* **86**, 4656 (2001).
54. Y. Takabayashi *et al.*, The Disorder-Free Non-BCS Superconductor Cs$_3$C$_{60}$ Emerges from an Antiferromagnetic Insulator Parent State. *Science* **323**, 1585 (2009).
55. Y. Okamoto, M. Nohara, H. Aruga-Katori, H. Takagi, Spin-Liquid State in the S=1/2 Hyperkagome Antiferromagnet Na$_4$Ir$_3$O$_8$. *Phys. Rev. Lett.* **99**, 137207 (2007).
56. S. Yamanaka, K. Hotehama, H. Kawaji, Superconductivity at 25.5K in electron-doped layered hafnium nitride. *Nature* **392**, 580 (1998).
57. H. Q. Yuan *et al.*, Observation of Two Distinct Superconducting Phases in CeCu$_2$Si$_2$. *Science* **302**, 2104 (2003).
58. G. Knebel, D. Aoki, J. Flouquet, Magnetism and Superconductivity in CeRhIn$_5$. http://arxiv.org/abs/0911.5223
59. A. Huxley *et al.*, Realignment of the flux-line lattice by a change in the symmetry of superconductivity in UPt$_3$. *Nature* **406**, 160 (2000).
60. J. D. Strand *et al.*, Evidence for Complex Superconducting Order Parameter Symmetry in the Low-Temperature Phase of UPt$_3$ from Josephson Interferometry. *Phys. Rev. Lett.* **103**, 197002 (2009).
61. M. R. Norman, D. Pines, C. Kallin, The pseudogap: friend or foe of high $T_c$? *Adv. Phys.* **54**, 715 (2005).
62. J. R. Kirtley *et al.*, Angle-resolved phase-sensitive determination of the in-plane gap symmetry in YBa$_2$Cu$_3$O$_{7-\delta}$. *Nat. Phys.* **2**, 190 (2006).
63. Y. Kurosaki, Y. Shimizu, K. Miyagawa, K. Kanoda, G. Saito, Mott Transition from a Spin Liquid to a Fermi Liquid in the Spin-Frustrated Organic Conductor κ-(ET)$_2$Cu$_2$(CN)$_3$. *Phys. Rev. Lett.* **95**, 177001 (2005).
64. S. Nandi *et al.*, Anomalous Suppression of the Orthorhombic Lattice Distortion in Superconducting Ba(Fe$_{1-x}$Co$_x$)$_2$As$_2$ Single Crystals. *Phys. Rev. Lett.* **104**, 057006 (2010).

65. This work was supported by the US DOE, Office of Science, under Contract No. DE-AC02-06CH11357 and by the Center for Emergent Superconductivity, an Energy Frontier Research Center funded by the US DOE, Office of Science, under Award No. DE-AC02-98CH1088.

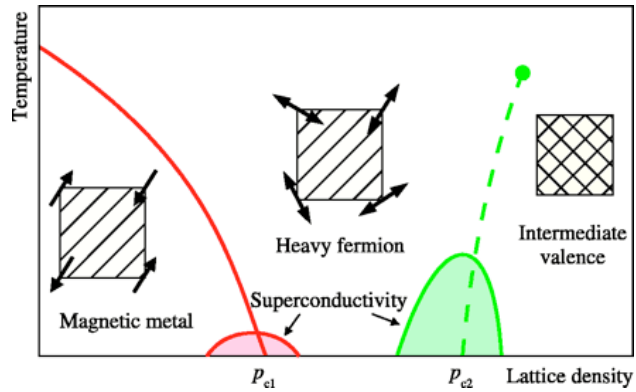 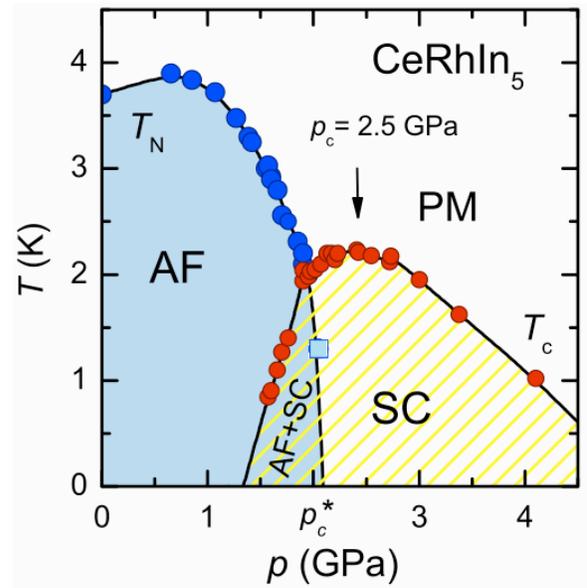

Figure 1 – **(A)** Schematic phase diagram of the heavy fermion material $CeCu_2Si_{2-x}Ge_x$. Two superconducting domes are present, one (red) associated with a quantum critical point of an antiferromagnet (at $p_{c1}$), the other (green) with a volume collapse transition (at $p_{c2}$) (*57*). **(B)** Phase diagram of $CeRhIn_5$ (temperature $T$ versus pressure $p$). The superconducting (SC) phase with a critical temperature $T_c$ abuts an antiferromagnetic (AF) phase with a Néel temperature $T_N$, with a small coexistence region in between and a paramagnetic (PM) phase at higher temperatures (*58*).

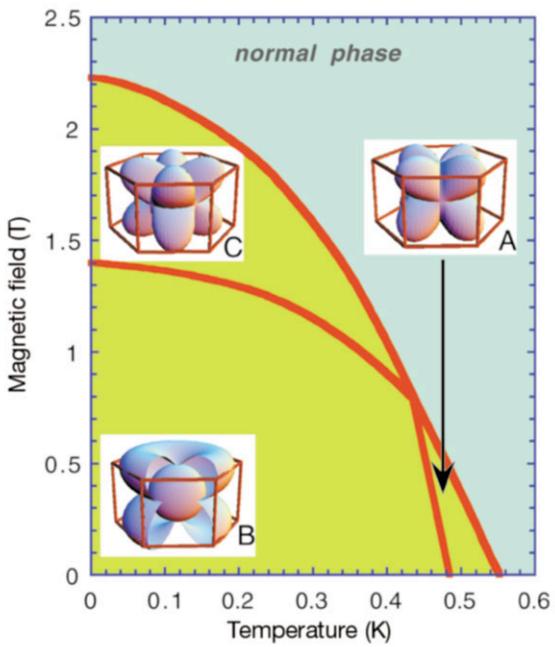 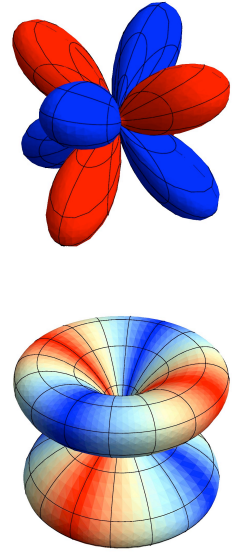

Figure 2 – **(A)** Phase diagram of UPt$_3$. Three different superconducting phases, *A, B* and *C*, are evident (*59*). **(B)** f-wave (E$_{2u}$) Cooper pair wavefunction proposed for phase *A* (top) and phase *B* (bottom) based on phase sensitive tunneling (*60*).

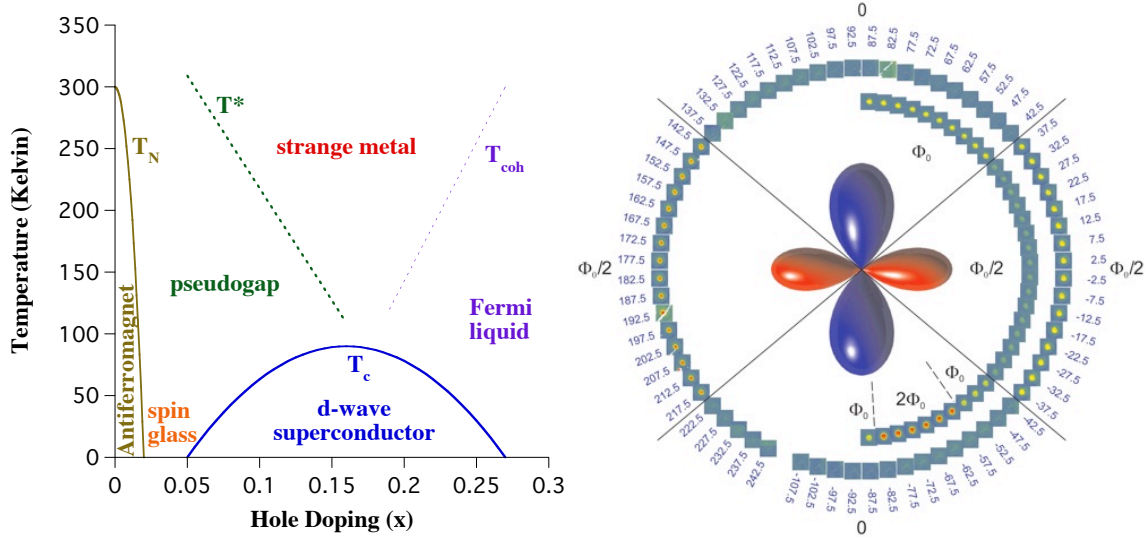

Figure 3 – **(A)** Schematic phase diagram of the cuprates (*61*). **(B)** Cooper pair wavefunction for $YBa_2Cu_3O_7$ as a function of the planar azimuthal angle measured by phase sensitive tunneling (*62*). The red and blue lobes denote the opposite signs of the $d_{x^2-y^2}$ state. The size difference of the two lobes is due to the orthorhombicity of this material, which leads to mixing in of a small s-wave component.

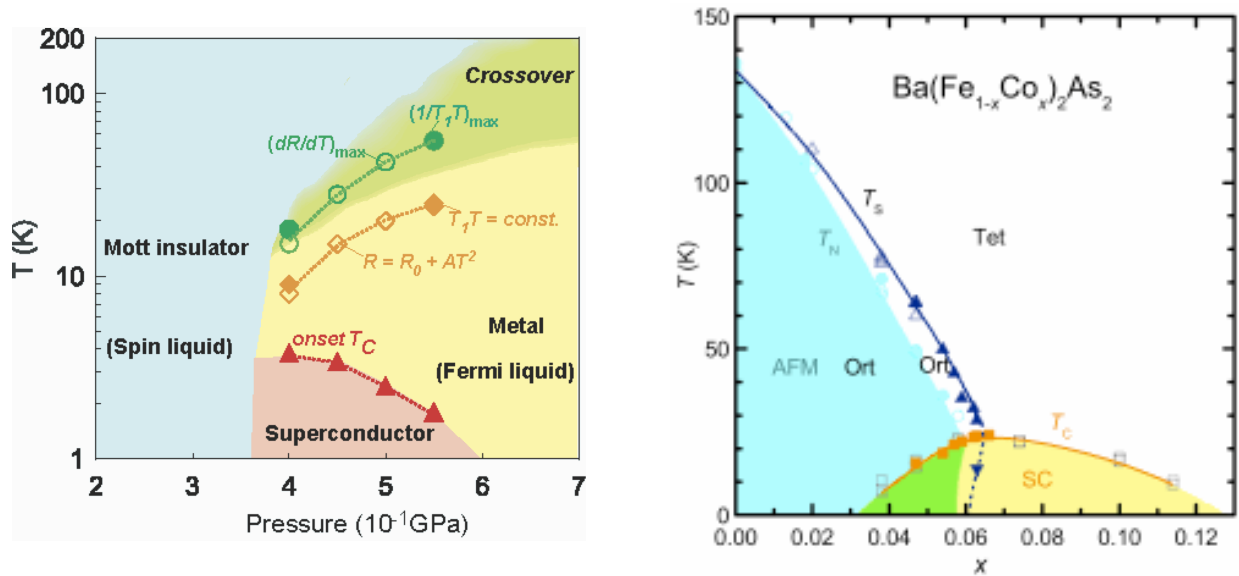

Figure 4 **(A)** Phase diagram of the organic superconductor κ-(ET)$_2$Cu$_2$(CN)$_3$ (temperature T versus pressure) (*63*). No magnetic order has been detected in the Mott insulator phase. **(B)** Phase diagram (temperature *T* versus doping *x*) of the pnictide superconductor Ba(Fe$_{1-x}$Co$_x$)$_2$As$_2$ (*64*). The antiferromagnetic (orthorhombic) phase occurs below T$_N$ (T$_s$) and is marked by AFM (Ort), the normal (tetragonal) phase by Tet, and the superconducting phase (below T$_c$) by SC. Note the similarities of the phase diagrams presented in Figures 1, 3, and 4.